\documentclass[]{aa}
\usepackage{times}
\usepackage{amssymb}
\usepackage{epsfig}
\def\pin{{\sc pinocchio}}
\def\mnras{MNRAS~}
\def\apj{ApJ}
\def\apjs{ApJS}

\def\msun{\hbox{${\rm M}_\odot$}}
\def\Mpc{\hbox{\rm Mpc}}

\title{Tracing large-scale structure at high redshift with Lyman-$\alpha$ 
emitters: the effect of peculiar velocities}

\author{P.~Monaco\inst{1} \and P.~M\o ller\inst{2} \and J.~P.~U.~Fynbo\inst{3,4} \and M.~Weidinger\inst{4} \and C.~Ledoux\inst{5} \and T.~Theuns\inst{6,7}}

\institute{Dipartimento di Astronomia, Universit\`a di Trieste,
           Via Tiepolo 11,
           34131 Trieste, Italy
\and
            European Southern Observatory,
            Karl-Schwarzschild-Stra\ss e 2,
            D-85748, Garching by M\"unchen,
            Germany
\and        
            Astronomical Observatory, University of Copenhagen, 
            Juliane Maries Vej 30,
            DK-2100 Copenhagen {\O},
            Denmark
\and  
            University of Aarhus, Ny Munkegade,
            DK-8000 \AA rhus C,
            Denmark
\and     
            European Southern Observatory,
            Alonso de C\'ordova 3107, Casilla 19001, 
            Vitacura, Santiago 19, Chile  
\and			
				Institute for Computational Cosmology,
				Department of Physics,
            University of Durham, 
            South Road, Durham DH1 3LE, UK
\and
				University of Antwerp,
				Campus Drie Eiken,
				Universiteitsplein 1,
				B-2610 Antwerp,
				Belgium
          }
\offprints{P. Monaco}
\mail{monaco@ts.astro.it}
\date{Received / Accepted }

\begin{document}
\abstract{ We investigate the effect of peculiar velocities on the
redshift space distribution of $z\ga2$ galaxies, and we focus in
particular on Ly$\alpha$ emitters. We generate catalogues of dark
matter (DM) halos and identify emitters with halos of the same
co-moving space density ($M(Ly{\alpha}\ emitters)\approx
3\times10^{11}\msun$).  We decompose the peculiar velocity field of
halos into streaming, gradient and random components, and compute and
analyse these as a function of scale. Streaming velocities are
determined by fluctuations on very large scales, strongly affected by
sample variance, but have a modest impact on the interpretation of
observations. Gradient velocities are the most important as they
distort structures in redshift space, changing the thickness and
orientation of sheets and filaments. Random velocities are typically
below or of the same order as the typical observational uncertainty on
the redshift. We discuss the importance of these effects for the
interpretation of data on the large-scale structure as traced by
Ly$\alpha$ emitters (or similar kinds of astrophysical high-redshift
objects), focusing on the induced errors in the viewing angles of
filaments.  We compare our predictions of velocity patterns
for Ly$\alpha$ emitters to observations and find 
that redshift clumping of Ly$\alpha$ emitters, as reported for instance in
the fields of high-redshift radio galaxies, does not allow to infer whether
an observed field is sampling an early galaxy overdensity.
 }

\titlerunning{The effect of peculiar velocities}
\maketitle

\section{Introduction}
It has become feasible to obtain accurate redshifts for large samples
of distant objects and produce 3-dimensional maps of the universe out
to redshifts 3 and beyond. This has allowed quantitative studies of
the large-scale structure of the distant Universe using Lyman-Break
Galaxies (LBGs, see, e.g., Adelberger et al.\ 1998; Miley et al.\
2004), Ly$\alpha$ emitters (Warren \& M\o ller 1996; Steidel et al.\
2000; M{\o}ller \& Fynbo 2001; Fynbo, M{\o}ller \& Thomsen 2001;
Shimasaku et al.\ 2003), extremely red objects (Daddi et al.\ 2003),
or radio galaxies (Pentericci et al.\ 2000; Venemans et al.\ 2002) as
tracers. These surveys are consistent with the galaxies tracing the
characteristic filamentary pattern, aptly called the \lq cosmic
web\rq\, in the dark matter, a generic feature of structure formation
in a cold dark matter dominated universe.

In such 3D maps the third dimension is given by redshift and therefore
they are deformed by the peculiar velocities of the galaxies. For example,
infall onto clusters introduces a characteristic \lq finger of god\rq\
pattern in the CfA redshift survey (Huchra et al., 1983). The peculiar
velocities can be thought of as the sum of three
components $(i)$ a streaming flow (driven by fluctuations on very
large scales), $(ii)$ a coherent component that distorts the
large-scale structure and $(iii)$ a small-scale noise term due to
highly non-linear motions. Peculiar motions may change the apparent
inclination angle at which filaments are observed, so if one wishes to
apply an Alcock-Paczy\'nski test on the distribution of inclination
angles of filaments (Alcock \& Paczy\'nski 1979; M{\o}ller \& Fynbo
2001), then it is imperative to understand at which scales such an
analysis will not be systematically biased (Weidinger et al.\
2002). Peculiar motions may also lead to an apparent enhancement in
the clumpiness of the redshift distribution of Ly$\alpha$ emitters in
deep narrow-band observations, as we will show below.

In this paper we follow Haehnelt, Natarajan \& Rees (1998) and
identify Ly$\alpha$ emitters with halos in our DM simulations by
requiring that the halos of the corresponding mass have the same
number density as the Ly$\alpha$ emitters. Such a simple biasing
scheme should work well if the duty cycle of Ly$\alpha$ emission is
close to one.  We then investigate to what extent peculiar velocities
influence the observed distribution of Ly$\alpha$ emitters. Although
we focus on the large-scale structure as traced by Ly$\alpha$
emitters, our conclusions can be applied to other classes of objects
as well.

This paper is organised as follows.  In Sect.~2 we decompose the
velocity field of DM halos into streaming, gradient and random
components, and show how to estimate such velocity components on DM
halo catalogues generated with the \pin\ code (Monaco et al. 2002).
In Sect.~3 we quantify the velocity components and give analytic fits
to reproduce the results.  The observational consequences of these
results on the reconstruction of viewing angles of filaments and on
the redshift distribution of Ly$\alpha$ emitters in narrow band
imaging selected volumes at high redshift are given in Section~4.
Section~5 concludes. More details on the extension of the \pin\ code
to multi-scale runs and on the connection between DM halos and
observed astrophysical objects are given in three appendices.

In this paper we assume a
scale invariant, vacuum energy dominated flat universe with parameters
$(\Omega_m,\Omega_\Lambda,n,h,\sigma_8)=(0.3,0.7,1,0.7,0.9)$ (e.g.
Spergel et al. 2003), where the symbols have their usual meaning.

\section{Quantification of peculiar velocity components}
\label{section:quantification}

Consider a set of DM halos in a given volume; the (highly correlated)
peculiar velocity field traced by these halos can be decomposed in
three components that have different effects on observations in the
redshift space (e.g. Weidinger et al. 2002): {\it (i)} the mean
velocity of the set, or streaming velocity, {\it (ii)} a gradient
component of velocities along the volume and {\it (iii)} the residual
(random) component. Performing a Taylor expansion of the peculiar
velocity around the set's mean velocity, these components are:

\begin{eqnarray}
{\bf x} &=& {\bf r}/a\nonumber\\ {\bf v}({\bf x}) &\equiv&a{d{\bf
x}\over dt}\nonumber\\ &=&{\bf v}({\bf x}_0) + \left[({\bf x}-{\bf
x}_0)\cdot \nabla_{\bf x}\right] {\bf v}\, ({\bf x}_0)+ {\bf
v}_{r}({\bf x}). \label{eq:def}
\end{eqnarray}

Here, $a$ is the scale factor, $\bf{x}$ is the co-moving position of
the halo, and the peculiar velocity ${\bf v}$ at position $\bf{x}$ is
written as a sum of a streaming, gradient and random component. The
geometrical meaning of the three terms is straighforward: the
streaming velocity gives the bulk motion of the whole set; the gradient
component is responsible for distortions of the redshift-space
patterns and can change the shape of the structure traced by the set;
random velocities are considered here as a noise.  Note that the
gradient component is a tensor, but we will restrict ourselves to the
diagonal components:

\begin{equation}
h_i=\partial v_i/\partial x_i\,.
\end{equation}

These velocity components are of course only defined once the volume
containing the set has been identified, {\em i.e.}
once a (co-moving) scale has been chosen.  In other words, these
quantities are scale-dependent.

\subsection{Velocity components in linear theory}
\label{section:linear theory}

\begin{figure*}
\centering{ 
\epsfig{file=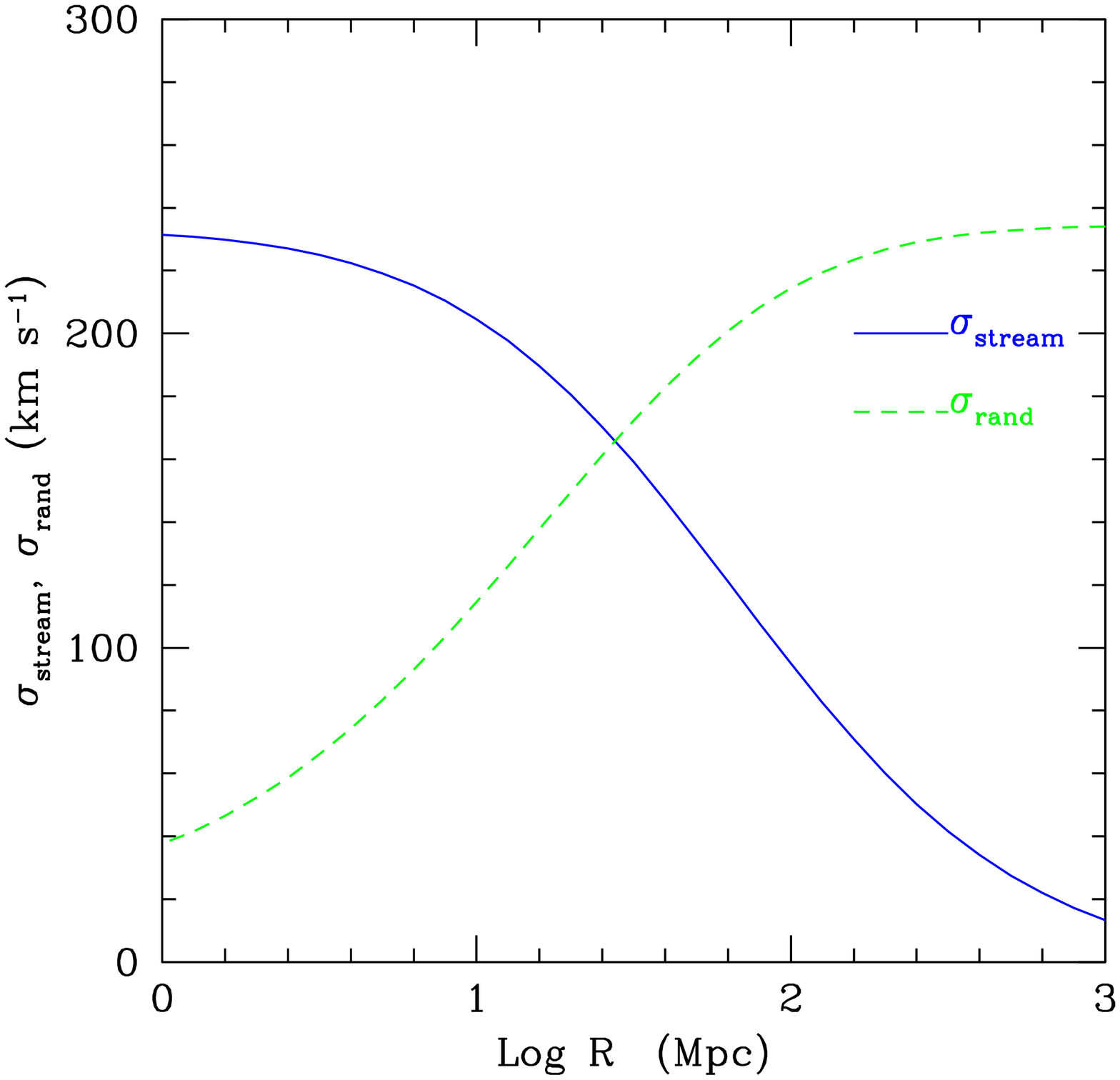,width=8.0cm}
\epsfig{file=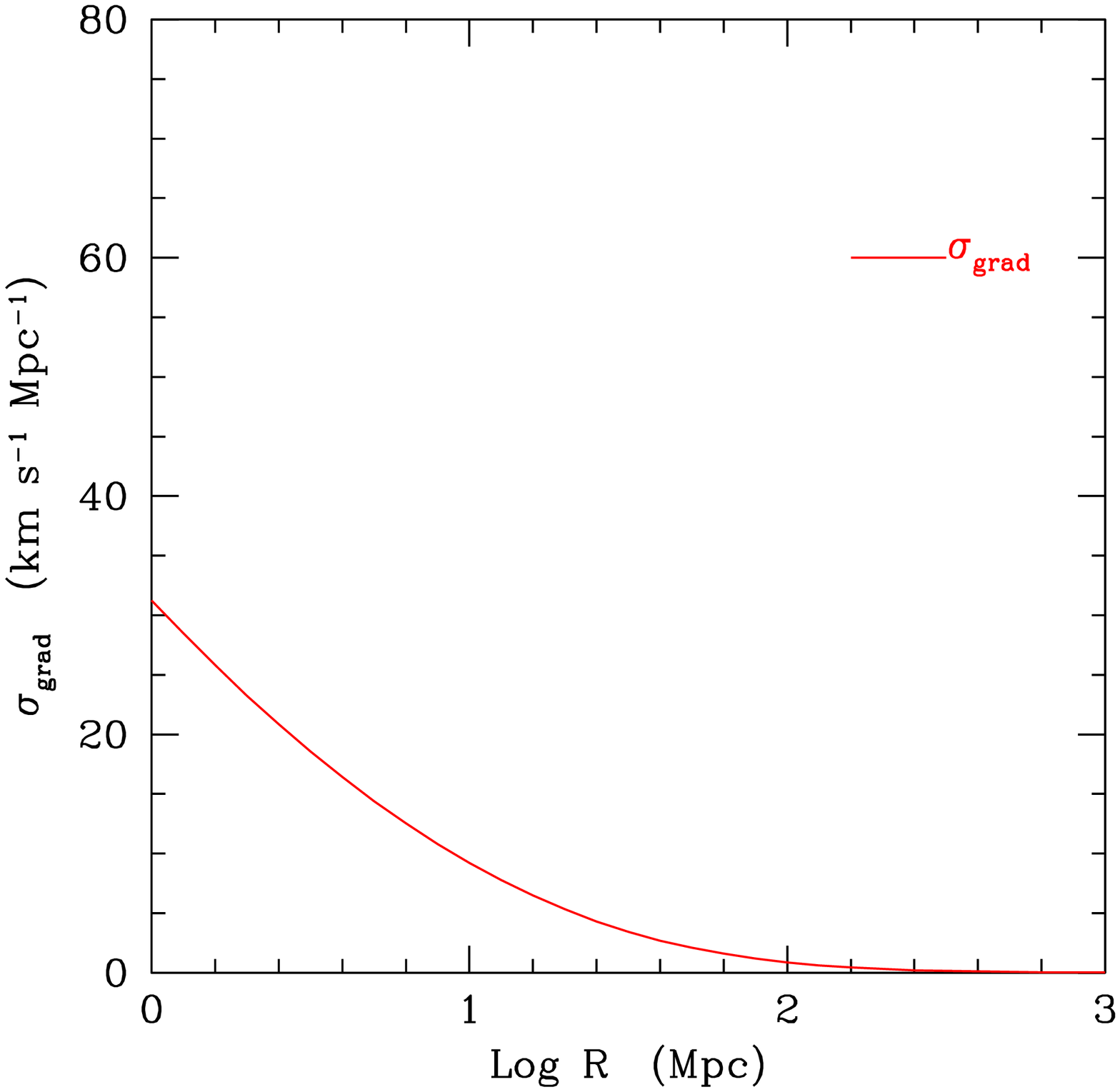,width=8.0cm} }
\caption{Predictions of linear theory for the three velocity
statistics.  Left panel: streaming and random velocities
(equations~\ref{eq:sigmabulk} and \ref{eq:sigmarand}).  Right panel:
gradient velocity (equation~\ref{eq:sigmagrad}).}
\label{fig:theo}
\end{figure*}

It is useful to consider the predictions of linear theory for the
three velocities defined above (equation~\ref{eq:def}) as a function
of scale, both to have an idea of their behaviour and as a basis for
constructing analytic fits of the simulation results in the next
section (see also Hamana et al. 2003).

In the Zel'dovich (1970) approximation, the peculiar velocity ${\bf v}$
of a halo is related to the gravitational potential $\phi$ by
(e.g. Peebles 1980)

\begin{equation}
{\bf v} = -a(t) \dot{D}(t) \nabla\phi\, \label{eq:pecvel}\,,
\end{equation}

\noindent
where a dot denotes a time derivative. In turn, $\phi$ is related to
the underlying density field $\delta({\bf x},t)\equiv \rho({\bf
x},t)/\langle \rho({\bf x},t)\rangle$ through Poisson's equation,

\begin{equation}
\nabla^2\phi({\bf x},t) = {1\over a}\,\delta({\bf x},t)\,.
\end{equation}

In these equations $D(t)$ 
is the
growing mode of the density perturbation. If $P(k)$ denotes the
power spectrum of density fluctuations, then Poisson's equation combined
with equation~(\ref{eq:pecvel}) shows that the spectrum of velocity
perturbations is $P_v\propto k^{-2}\,P(k)$. Therefore the (1D) variance
of the peculiar velocity in linear theory is

\begin{equation}
\sigma^2_v = \frac{1}{3} (a\dot{D})^2 
\frac{1}{2\pi^2}\int_0^\infty P(k)dk\,.
\label{eq:sigmav} \end{equation}

\noindent
Note that $\sigma^2_v$ converges readily on small scales where
$P(k)\propto k^n$ with $n\sim -2$ to $-3$, but converges only slowly
on large scales, as is well known.

We now focus on a system of size $R$ to decompose $\sigma^2_v$ in terms
of a streaming, gradient, and random component. The variance of the
streaming velocity of the matter in the volume $V$ can be computed by
smoothing $\phi$ with a window function $W_R({\bf q})$ (with $V\sim
R^3$). If $\tilde{W}$ denotes the Fourier transform of $W$, then

\begin{eqnarray}
\sigma^2_{\rm stream}(R) &=& \frac{1}{3} (a\dot{D})^2 \sigma^2_0(R)
\label{eq:sigmarand}\\
\sigma^2_{\rm rand}(R) &=& \sigma^2_v - \sigma^2_{\rm stream}(R),
\label{eq:sigmabulk}
\end{eqnarray}

where 

\begin{equation}
\sigma^2_l(R)\equiv\frac{1}{2\pi^2}
\int_0^\infty k^l\,P(k)\tilde{W}^2(kR) dk \, . \label{eq:sigmal}
\end{equation}

Similarly, the rms gradient

\begin{equation}
\sigma^2_{\rm grad}(R) = \frac{1}{3} (a\dot{D})^2\,\sigma^2_2(R)\,.
\label{eq:sigmagrad}
\end{equation}

\noindent
Note that $\sigma^2_2(R)$ in this equations is also the mass variance
on scale $R$.  In the following we will use a top-hat filter, so that
the mass variance will be related to spherical volumes of radius $R$.

In Fig.~\ref{fig:theo} we plot the streaming, gradient and random
components predicted by linear theory at $z=2$ as a function of scale
$R$.  Streaming velocities dominate on scales smaller than $\sim30$
Mpc; on larger scales random velocities dominate.  The cross-over
scale is a measure of the velocity correlation in linear theory, and
is similar to the size of the largest observable structure (filament
or void) at the given redshift. The rms gradient $\sigma_{\rm
grad}(R)$ is usually a small fraction of the Hubble constant $H(z)$.

In the next section we compute these variances using simulations.

\subsection{Simulations}
\label{section:simulations}

As we saw in the previous section, peculiar velocity fields are
correlated over large scales, hence simulations need to be performed
in a large volume to adequately sample the large-scale modes. Combined
with the need to be able to resolve small halos, we require
simulations with a large dynamic range. Furthermore, to test
convergence of results and decrease sample variance we need to run
many simulations. The \pin\ algorithm (Monaco et al.\ 2002a; Monaco,
Theuns \& Taffoni 2002b; Taffoni, Monaco \& Theuns 2002) is ideally
suited for this purpose.

\pin\ uses Lagrangian perturbation theory and an algorithm to mimic
the hierarchical build-up of DM halos to predict the masses, positions
and velocities of dark matter halos as a function of time. The
agreement between \pin\ and a full scale $N$-body simulation is very
good, even when comparing the properties of individual halos. \pin\
does not compute the density profile of the halos, and as a
consequence is many orders of magnitude faster than an $N$-body
simulation. A simulation with 256$^3$ particles requires a few hours
on a PC.

As we will show below, a typical Ly$\alpha$ emitter lies in a halo
of mass $\sim 3\times 10^{11}\msun$, so if we want to resolve a halo
with 150 particles, then the particle mass is $2\times 10^9\msun$. In a
simulation with $256^3$ particles, and given our assumed cosmology,
this limits our box size to $\sim 65\, h^{-1}\Mpc$, too small to properly
sample the large-scale velocity field. Fortunately it not necessary to
perform much larger, computationally expensive simulations, because it
is straightforward to add long-wavelength perturbations to \pin.  This
is explained in Appendix A, while Appendix B quantifies the accuracy
of \pin\ in reproducing the velocity components defined above.

To address convergence of peculiar velocities and sample variance, we
have run many realisations (Table \ref{tab:runs}).  For the reference
cosmology, we have performed 10 standard \pin\, runs with a single
grid of size $L=65\, h^{-1}$ Mpc and grid spacing $\Delta$ such that
$L/\Delta=256$ ($\Delta=0.254\, h^{-1}$ Mpc), and 11 \pin\, runs with
two grids, using a high-resolution grid with $L_2=65\, h^{-1}$,
$L_2/\Delta_2=256$, and a low-resolution with $L_1= 8 L_2=520\,
h^{-1}$ Mpc, $L_1/\Delta_1=64$ ($\Delta=8.125\, h^{-1}$ Mpc).

\begin{table}
\caption[]{\pin\, runs performed, the particle mass is $6.7\times
10^8$. $L_1$ and $\Delta_1$ ($L_2$ and $\Delta_2$) are the size of the
low (high) resolution box and grid spacing, respectively. Runs of given
type only differ in the random seed.}
\begin{tabular}{c c c c c c}
\hline
\hline
run id & \# of runs  & $L_1$ & $L_1/\Delta_1$ & $L_2$ &
$L_2/\Delta_2$ \\
      &                           & $h^{-1}\Mpc$ &  & $h^{-1}\Mpc$ \\

\hline
P1    & 10 & -   & -  & 65 & 256 \\
P2    & 11 & 520 & 64 & 65 & 256 \\
\hline
\end{tabular} 
\label{tab:runs}
\end{table}

\subsection{Computing velocity statistics from halo catalogues}
\label{section:statistics}

\begin{figure*}
\centering{
\epsfig{file=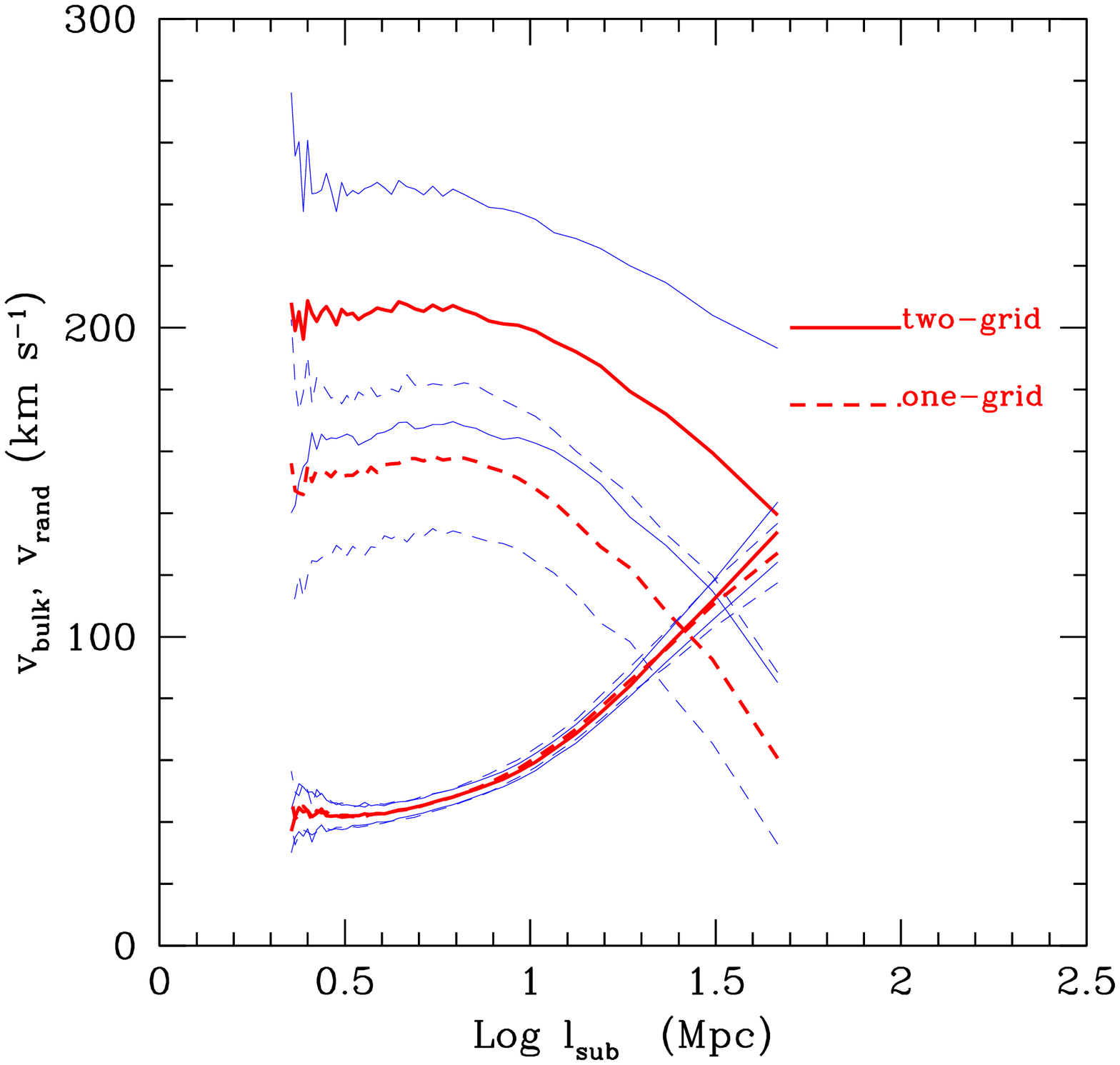,width=8.0cm}
\epsfig{file=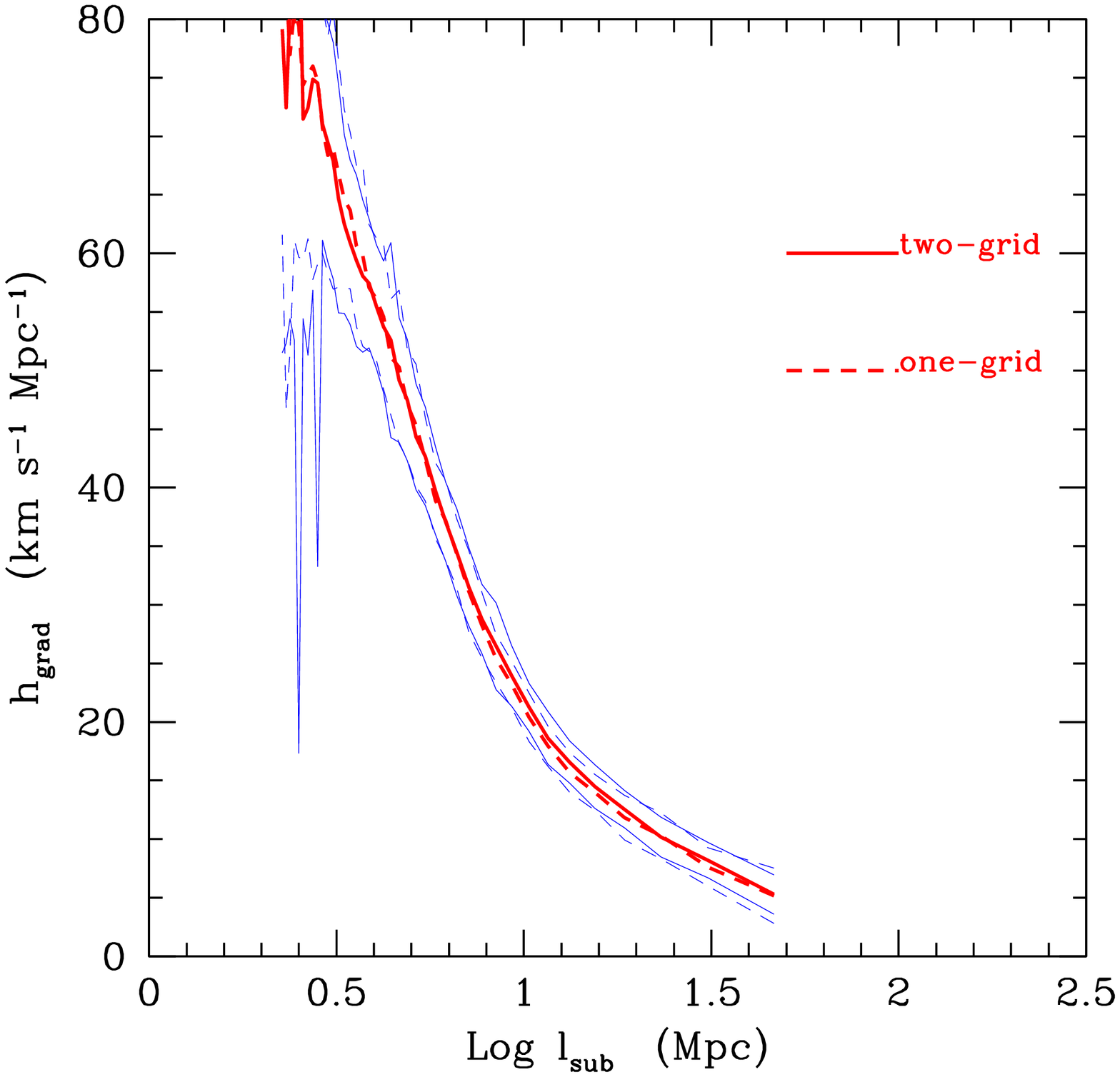,width=8.0cm}
}
\caption{Velocity statistics for halos with mass $M>10^{11}$ M$_\odot$
at $z=2$ in the \pin\, simulations of Table~1. Thick lines and thin
lines denote the mean and 1-$\sigma$ dispersion of these statistics
respectively, for runs P1 (dashed) and P2 (full lines). Note how the P1
simulation severely underestimates the streaming velocity because of
the missing large-scale power in the small simulation box.}
\label{fig:conv}
\end{figure*}

To compute streaming, gradient and random velocities from the
\pin\ runs, we have subdivided the simulated boxes into $n^3$
cubic sub-boxes of side $l_{\rm sub}=l_{\rm box}/n$, where $n$ is a
running integer.  For each subdivision $n$, each sub-box $(j,n)$
(where $j=[0,n^3]$) is centred on ${\bf x}_{0,j,n}$ and contains
$N_{j,n}$ dark matter halos more massive than a given threshold mass.
For each sub-box $(j,n)$ and for each spatial component $i$ we compute
the streaming and gradient velocities as the zero point and slope of a
linear regression with respect to position $x_i$ of the velocities
$v_i$ of all the halos:

\begin{equation}
v_i ({\bf x}_{0,j,n})
= \frac{\sum x_i^2 \sum v_i - \sum x_i \sum x_iv_i}
{N_{j,n}\sum x_i^2 - (\sum x_i)^2}
\label{eq:bulk} \end{equation}
\begin{equation}
\frac{\partial v_i}{\partial x_i} ({\bf x}_{0,j,n}) = 
\frac{N_{j,n}\sum x_iv_i - \sum x_i \sum v_i}
{N_{j,n}\sum x_i^2 - (\sum x_i)^2}
\label{eq:gradient} \end{equation}

\noindent
The sum is over the $N_{j,n}$ halos in the sub-box $(j,n)$.  
Random velocities are computed as the residuals of velocities with
respect to the linear regression.  For each sub-box $(j,n)$ and for
each component $i$ we compute the variance of these residuals:

\begin{eqnarray} \label{eq:random} 
\lefteqn{\sigma_{r,i}^2 ({\bf x}_{0,j,n}) =}
\\ &&{\rm var}\left(v_i({\bf
x})-v_i({\bf x}_{0,j,n}) -({\bf x}-{\bf x}_{0,j,n})_i \frac{\partial
v_i}{\partial{x_i}}({\bf x}_{0,j,n})\right) \nonumber
\end{eqnarray}

Finally, for each sub-box size $n$ we compute the variance of these
quantities over all sub-boxes $j$ that contain at least 5 objects
($N_{j,n}\ge 5$), and express the result as a function of the sub-box
length $l_{\rm sub}$:

\begin{eqnarray}
\nonumber
v_{{\rm stream},i}^2 (l_{\rm sub}) & = & \left\langle  \left( 
v_i ({\bf x}_{0,j,n})
\right)^2 \right\rangle_j \\
\label{eq:defs}
h_{{\rm grad},i}^2 (l_{\rm sub}) & = & \left\langle \left(
\frac{\partial v_i}{\partial x_i} ({\bf x}_{0,j,n})
 \right)^2 \right\rangle_j \\
\nonumber
v_{{\rm rand},i}^2 (l_{\rm sub}) & = & \left\langle 
\sigma^2_{r,i}  ({\bf x}_{0,j,n})
 \right\rangle_j
\end{eqnarray}

Notice that the second quantity, $h_{{\rm grad},i}$, has the same
dimension as the Hubble constant, while the other two are pure
velocities.  In the following, where not otherwise stated, we will
show for the three velocity statistics the average over the three
directions.

\section{Results}
\label{section:results}

Fig.~\ref{fig:conv} shows the streaming, gradient and random velocity
variances for halos larger than $10^{11}$ M$_\odot$ at $z=2$, as a
function of scale averaged over the P1 simulations (dashed lines).  As
for linear theory, on small scales streaming flows dominate and random
velocities are relatively small, while the reverse is true at larger
scales.  However, the relatively modest value of the streaming
velocity is mostly an artifact connected to the small box size and the
resulting poor sampling of the large scale modes; in a low-density
universe velocity fields are highly correlated, and to achieve
convergence for the streaming velocity it is necessary to average over
volumes as large as several hundred Mpc.  Fig.~\ref{fig:theo} shows
that the expected streaming flow velocity drops below 50 km s$^{-1}$
only at scales of $\sim$200 Mpc. The P2 simulations do not suffer from
lack of large scale power, and the streaming velocities are much
larger (Fig.~\ref{fig:conv}). Note that the random and gradient
dispersions are similar in the smaller boxes.

\begin{figure}
\centering{
\epsfig{file=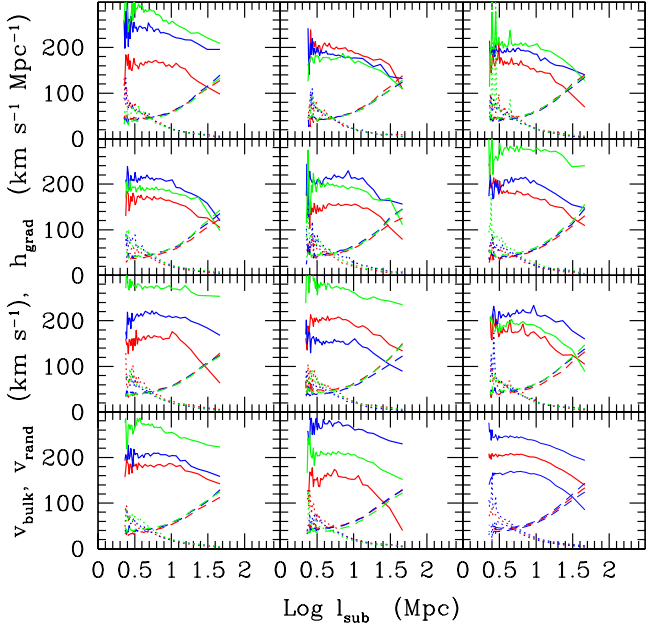,width=9.0cm}
}
\caption{Sample variance of velocity statistics for the 11 runs of the
P2 simulation.  Continuous, dotted and dashed lines denote streaming,
gradient and random velocity variances, respectively.  Here we
show the three spatial components separately.  The 12th panel contains
the average variance for all runs, with the corresponding 1-$\sigma$
dispersion.}
\label{fig:samplevar}
\end{figure}

Due to the large correlation length of the velocity field, the three
velocity statistics (especially the streaming motions) show a great
deal of variation among different realisations, which we call sample
variance.  Fig.~\ref{fig:samplevar} shows the random and streaming
velocities for each of the P2 runs separately.  Notice that the P1
realisations artificially suppress some of the sample variance by
imposing periodic boundary conditions, while the P2 realisations do not
have this drawback.  Here we show in each panel the streaming, gradient
and random velocities as computed in the three directions.  The curves
fluctuate much from one realisation to the other.  The 12th panel shows
the average with 1$\sigma$ fluctuations, the same quantity shown in
Fig.~\ref{fig:conv}.

This figure shows clearly the importance of a proper quantification of
sample variance.  This is important not only to test the reliability
of the predictions, but also to quantify the interval in which
observed data are expected.

\begin{figure*}
\centering{
\epsfig{file=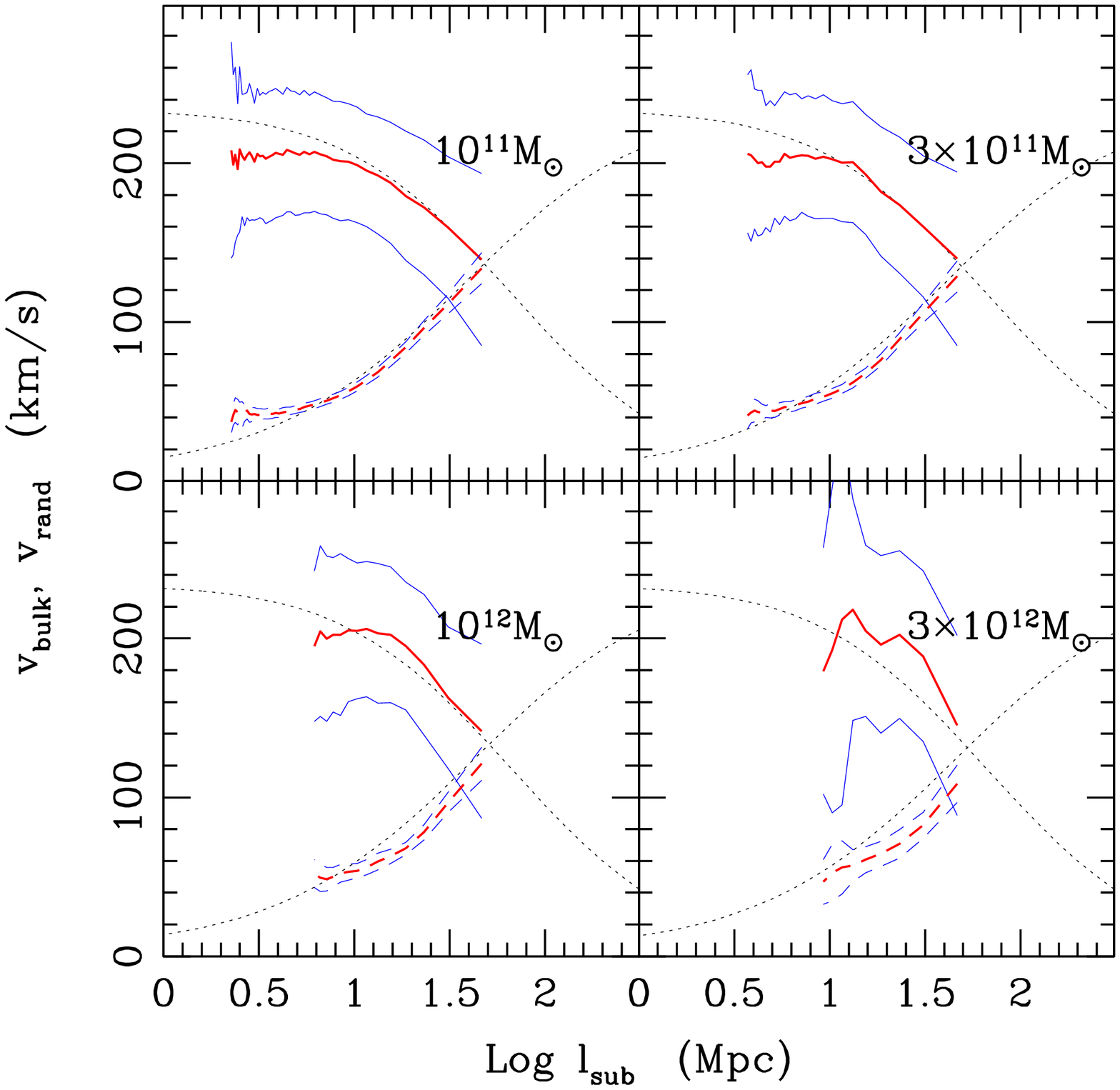,width=8.5cm}
\epsfig{file=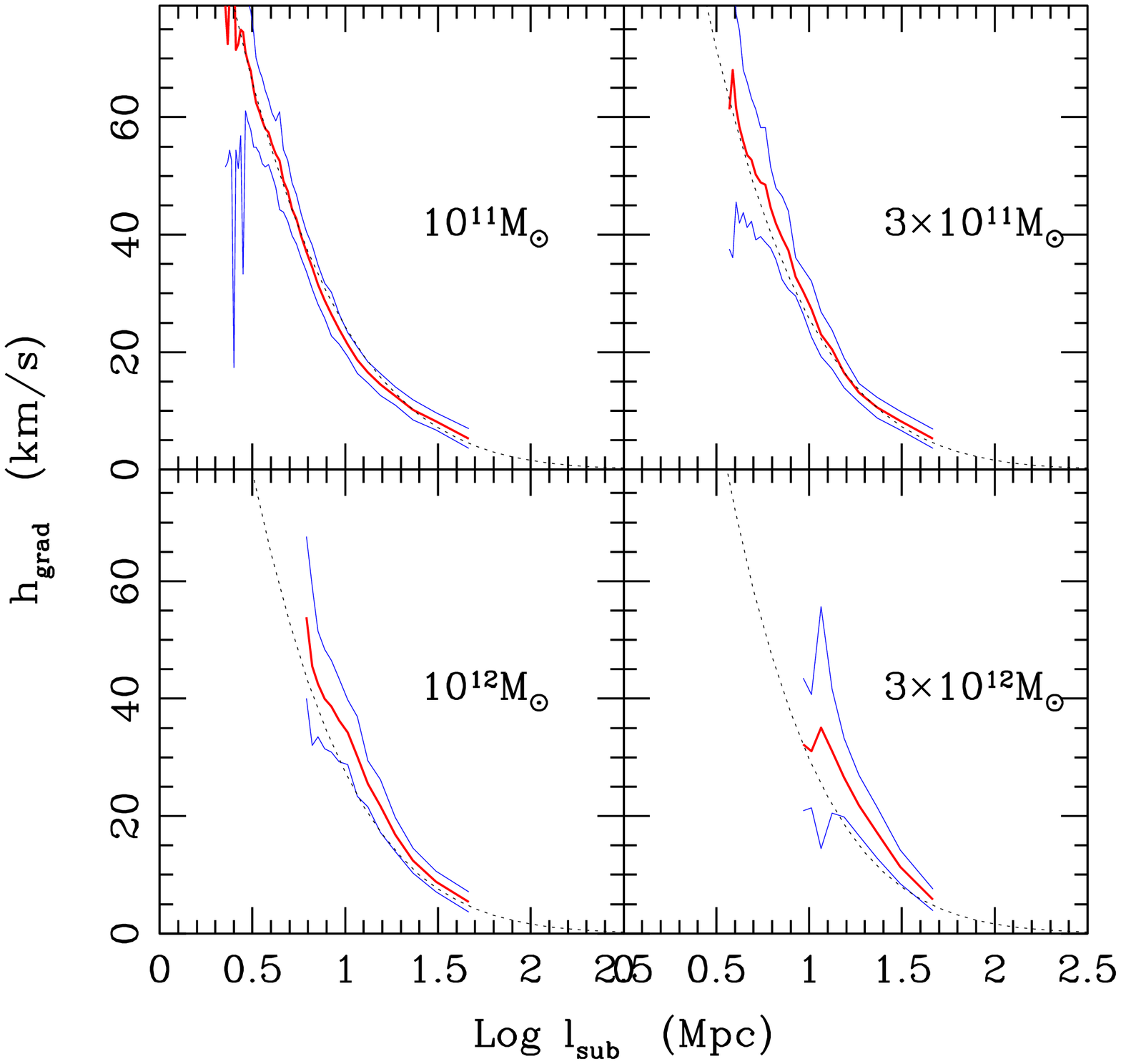,width=8.5cm}}
\caption{Velocity statistics for different halo masses at $z=2$.  The
average and variance for the simulation P2 are shown.  Dotted lines
give the analytic fits (equations~\ref{eq:sigmabulk}, \ref{eq:gradfit}
and \ref{eq:randfit}).}
\label{fig:mass}
\end{figure*}

\begin{figure*}
\centering{
\epsfig{file=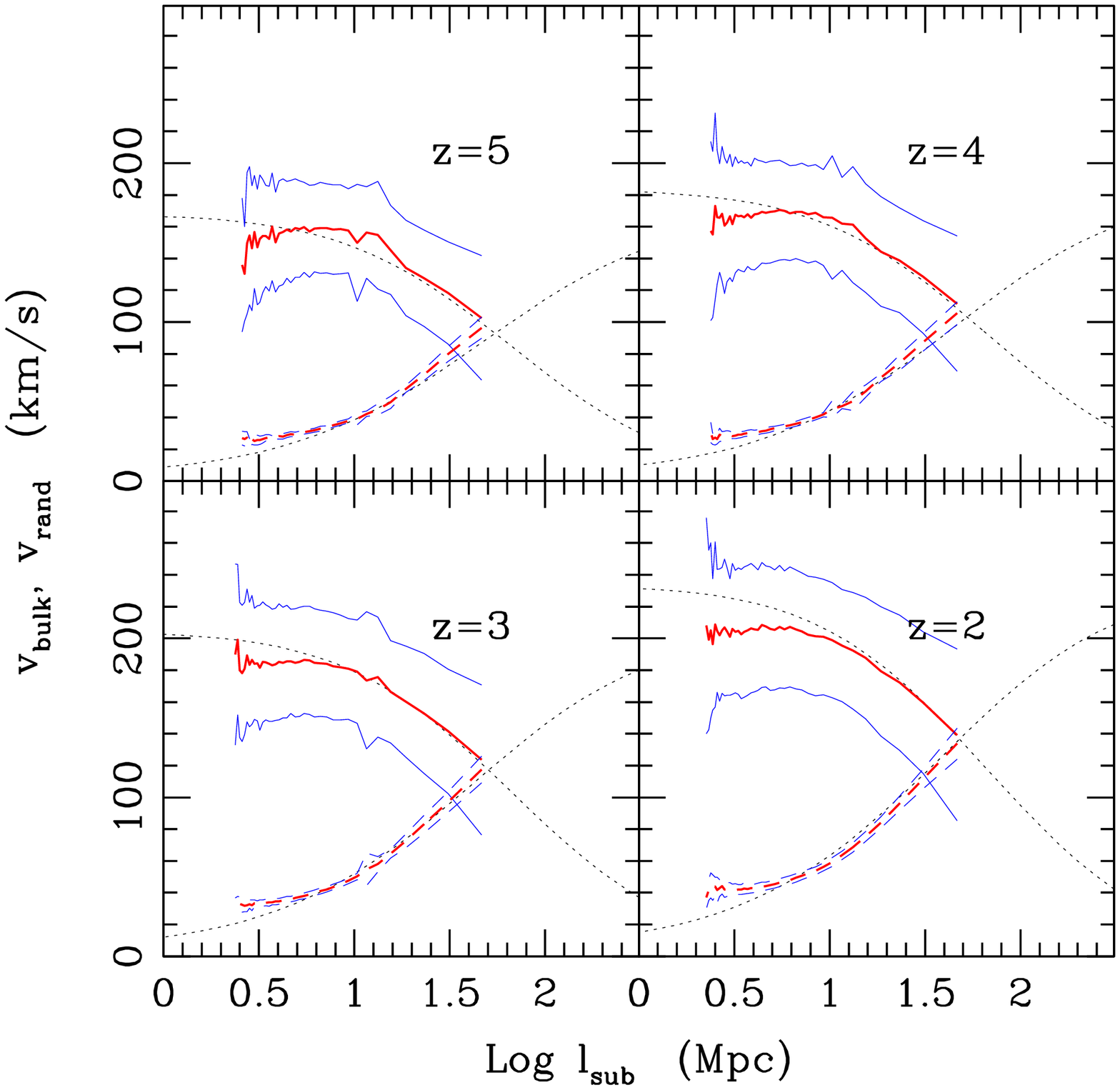,width=8.5cm}
\epsfig{file=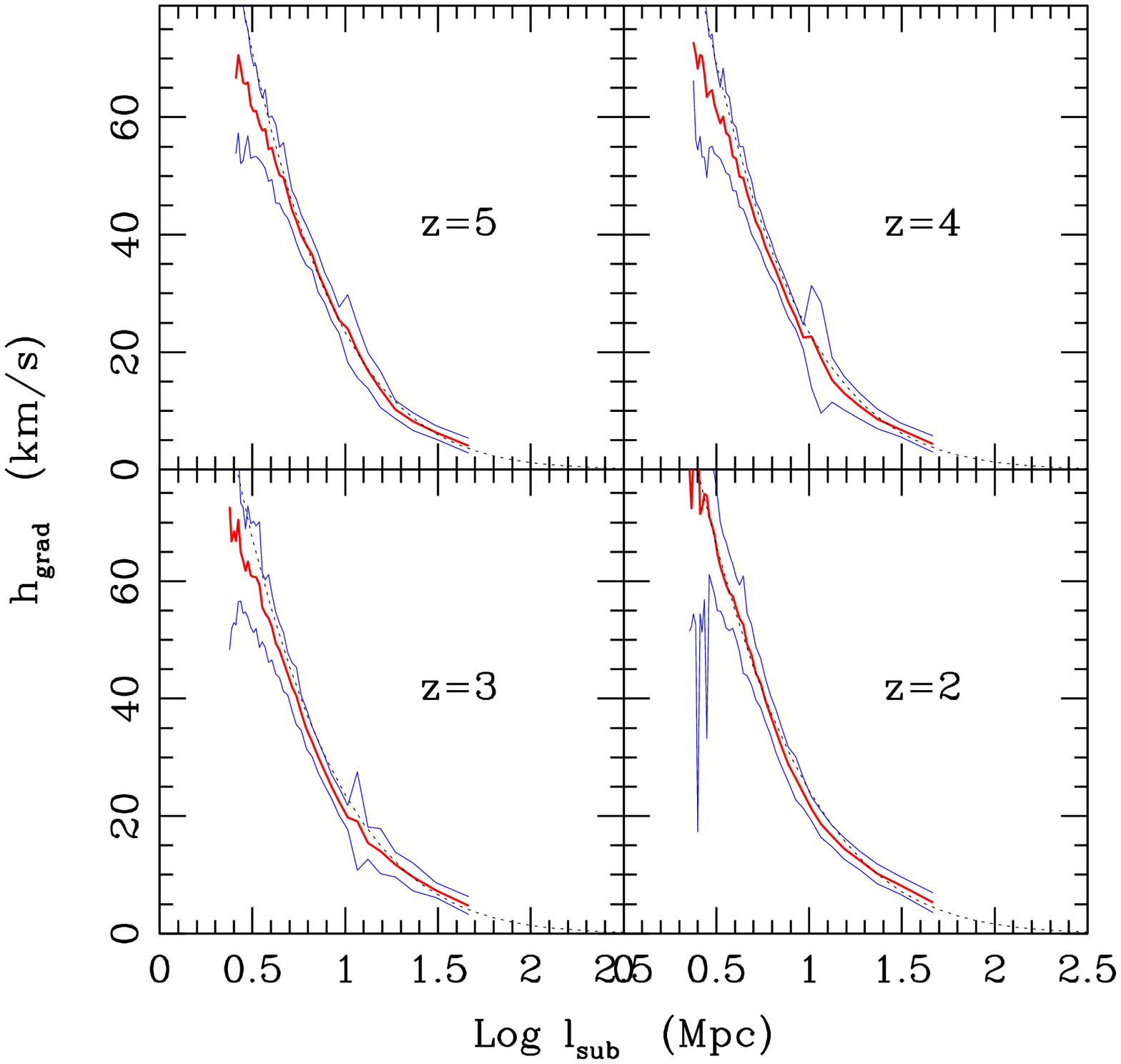,width=8.5cm}}
\caption{Same as Fig.~\ref{fig:mass} but for $M>10^{11}$ M$_\odot$ and
different redshifts.}
\label{fig:red}
\end{figure*}

The velocity statistics shown above depend on DM halo mass, redshift
and cosmology.  Figs.~\ref{fig:mass} and \ref{fig:red} illustrate the
mass and redshift dependence. In the same figures we show analytic
fits, based on linear theory, to the velocity statistics.  In
particular, streaming velocities are reasonably well fit on the scales
of interest by the simple linear theory prediction of
equation~\ref{eq:sigmabulk}.  A weak mass dependence is noticeable,
however it is much stronger for gradient and random velocities.  Such a
mass dependence is expected since halos are biased tracers of the
mass (see also Hamana et al. 2003).

Because of the self-similar character of gravity, we expect to be able
to fit the mass-dependence by a simple function of the spectral moments
$\sigma_l$ (equation~\ref{eq:sigmal}).  The mass of the halo depends on
$\sigma_2(R)$, while the velocity variances depend on $\sigma_0(R)$
(equations~\ref{eq:sigmarand} and \ref{eq:sigmabulk}).  The (top-hat)
co-moving smoothing radius $R$ is connected to the halo mass $M$ through
the relation $4\pi R^3\rho_0 =M$, where $\rho_0$ is the actual average
matter density.  With this $M-R$ relation the mass variance relative to
$M$, $D(z)\sigma(M)$ is then computed\footnote{Notice that the mass
variance $\sigma(M)$ is, by definition, linearly extrapolated to $z=0$,
to obtain its value at $z$ it is multiplied by the growing mode
$D(z)$.}.  The mass dependence of random and gradient velocities is
then reasonably well reproduced (Fig.~\ref{fig:mass}, \ref{fig:red}) by

\begin{equation}
h_{\rm grad,fit}(R) = \frac{1}{\sqrt{3}} (a\dot{D}) \sigma_2(R)
\left(1+0.8\frac{\sigma}{D(z)\sigma_2(R)}\right)
\label{eq:gradfit}
\end{equation}

\begin{equation}
v_{\rm rand,fit}(R) = \frac{1}{\sqrt{3}} (a\dot{D}) \sigma_0(R)
\left[1+\left(\frac{\sigma}{D(z)\sigma_2(R)}\right)^{0.5}\right]^{-1}
\label{eq:randfit}
\end{equation}

\noindent
where $4\pi R^3\rho_0 =M$. They give acceptable fits at scales larger
than 10 co-moving Mpc, although some residual mass dependence is
present; in particular, more massive objects are not perfectly
reproduced.

We have verified that the dependence on cosmological parameters is
correctly reproduced by these fits by performing additional \pin\,
simulations, using the same random seeds to be less affected by sample
variance.

\section{Observational consequences}
\label{section:obs}

In the previous sections we characterised the effect of peculiar
velocities on the distribution of halos in redshift space. To apply
this result to galaxies we need to known how to associate galaxies
with dark matter halos. In this section we apply a very simple biasing
scheme where we associate galaxies of a given type with halos with the
same co-moving space density. More complex schemes have appeared
in the literature (e.g. based on the halo occupation distribution,
Berlind et al. 2003) but our model has the advantage of simplicity and
it is sufficient for our purpose\footnote{Notice that in this scheme
we only assume that a galaxy of a given (stellar) mass is typically
associated to a DMH of a given mass. We do not assume a direct
proportionality between galaxy and DMH mass, which we know it is not
present in real galaxies (see, e.g., Persic, Salucci \& Stel 1996).}.

Ly$\alpha$ emitters are the most numerous emission selected objects
known at high redshifts, suggesting that they must inhabit relatively
low-mass halos. In a deep search in two fields at z=2.85 and z=3.15,
Fynbo et al.\ (2003) determined the co-moving space density
$n_{Ly\alpha}$ of spectroscopically confirmed Ly$\alpha$ emitters down
to their Ly$\alpha$ flux detection limit of $7\times 10^{-18}$ erg
s$^{-1}$ cm$^{-2}$ to be $\log(n_{Ly\alpha})=-2.6$.  Assuming that
100\% of DM halos host a Ly$\alpha$ emitter, the measured space density
in our cosmology is typical of halos of mass $6\times10^{11}$
M$_\odot$.  The duty cycle could be lower than 100\%; for LBGs, only
25\% show significant Ly$\alpha$ emission (e.g., Shapley et al.\
2003). However, this is likely to be a lower limit to the duty cycle
of typical Ly$\alpha$ emitters, that have smaller star formation rates
and then are less affected by dust obscuration.  If a 25\% duty cycle
is adopted, the corresponding halo mass decreases to $2\times10^{11}$
M$_\odot$ .  These numbers should bracket the solution, and justify
the choice of $\sim3\times10^{11}$ M$_\odot$ anticipated in
Section~\ref{section:simulations}.

\subsection{The influence of velocities on Ly$\alpha$ filaments}

Several properties combine to make Ly$\alpha$ emitters a good tracer
for mapping large-scale structure. Because they have higher space
density than any other class of detectable objects at high redshifts
they provide the best possible sampling of structures at all scales,
their redshift is always measured from the same emission feature so
redshifts are obtained in a very homogeneous way, and their low masses
make them weakly biased tracers of the large-scale structure.  A
natural prediction of hierarchical clustering is then the likely
detection of filaments and pancakes in the 3D distribution of
Ly$\alpha$ emitters. One such filament traced by Ly$\alpha$ emitters
has been detected at $z=3.04$ (M\o ller \& Fynbo 2001), but the
inferred 3D properties of filaments will be modified by peculiar
velocities and to recover their true properties it is necessary to
understand those effects that can be divided into three distinct
components.

The streaming velocity of galaxies on the observed scale of the
filament will change the mean redshift by a small amount, $\sim 150$
km s$^{-1}$ on scales of tens of Mpc, amounting to a negligible shift
in redshift of $5\times 10^{-4}$. The gradient component will distort
the viewing angle of the filament; in particular the relative
(systematic) error on the line-of-sight dimension of the filament will
be:

\begin{equation}
\epsilon = \frac{\frac{\delta v}{\delta x}(1+z)}{H(z)}
\end{equation}

\noindent
(The gradient is multiplied by $(1+z)$ because the Hubble constant is
defined in terms of physical distance, in place of co-moving).  At that
scale the gradient will be of about 10 km $s^{-1}$ Mpc$^{-1}$, and the
relative error will be 0.13 (for a Hubble constant of 312 km $s^{-1}$
Mpc$^{-1}$, which is the Hubble constant at $z=3$ in the assumed
cosmology). This will also be the relative error of the arc cosine of
the viewing angle. The corresponding systematic error on the
inclination angle will hence typically be about 2-3$^o$, which is
similar to the 1.9$^o$ error due to sparse sampling on the inclination
angle of the $z=3.04$ filament (Weidinger et al.\ 2002).

Random velocities will thicken the filament.  For our test case,
velocities just above 100 km s$^{-1}$ are expected, so they will
contribute in a similar way as the typical uncertainty in the
redshift.

These effects should be taken into account when estimating, for
instance, the cosmological parameters by applying the extended
Alcock-Paczy\'nski test on the distribution of viewing angles
(M{\o}ller \& Fynbo 2001; Weidinger et al.\ 2002).

\subsection{Enhancement of clustering in redshift space}

The power of the approach presented here goes beyond a statistical
quantification of the effects of the velocity components.  We
illustrate this point by giving an example of interpretation of data
based on simulated catalogues of Ly$\alpha$ emitters.

Fynbo et al.\ (2003) detected a significant degree of redshift
clumping in the field around a DLA toward the quasar Q2138-4427 (at
$z=2.85$). This is clearly visible in their Fig.~8, where redshifts
clump into a limited interval, much narrower than the redshift-depth
corresponding to the filter. In the other field of that study
(Q1346-0322 at $z=3.15$), the redshifts are uniformly distributed over
the range defined by the filter.  The clumping can be quantified by
$\sigma_z$, the root-mean-square of the redshift distribution, found
to be 0.018 (with 19 emitters) and 0.006 (with 23 emitters) for the
fields of Q1346-0322 and Q2138-4427 respectively. These $\sigma_z$
values should be compared to the expected value of 0.019 based on a
simple Monte Carlo simulation using the filter transmission as
selection function. Hence, the Q2138-4427 field clearly shows a
significant degree of structure. Similar redshift clumping has been
reported in the fields of two radio galaxies at redshifts $z=2.14$ and
$z=4.10$ (Pentericci et al.\ 2000; Venemans et al.\ 2002).

It is interesting to ask how often and under what conditions does
similar redshift clumping occur in the simulations?  Peculiar
velocities can influence the clumping of redshifts in different ways.
While streaming flows shift the whole redshift distribution, gradient
velocities can increase or decrease the dispersion $\sigma_z$.  If a
mildly non-linear structure (a filament or a pancake) is present in
the field, it is known that the peculiar velocity field (its gradient
component, in our terminology) will tend to flatten it, thus decreasing
$\sigma_z$ (see, e.g., Strauss \& Willick 1995). Random velocities
will instead tend to increase $\sigma_z$.

\begin{figure}
\centering{
\epsfig{file=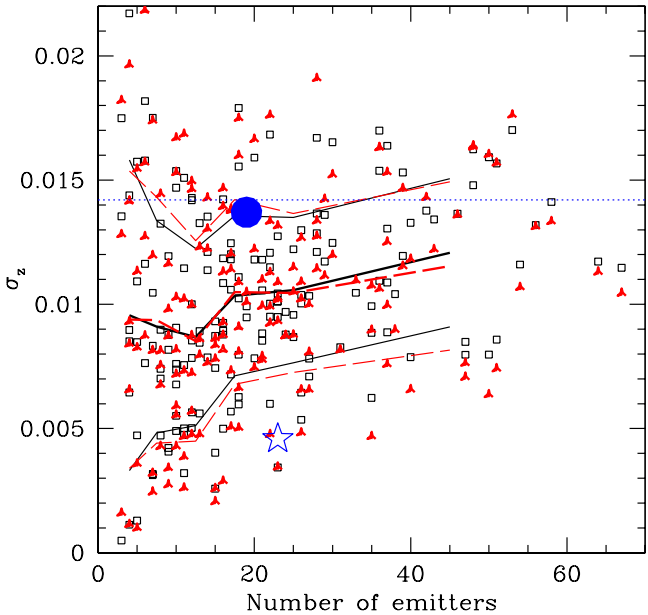,width=9.0cm}}
\caption{Redshift dispersion, $\sigma_z$, of Ly$\alpha$ emitters
selected in a narrow-band field as function of the number of
emitters. A big filled circle and a big star denote the fields around
Q1346-0322 and Q2138-4427 (Fynbo et al. 2003), respectively. Filled
triangles denote $\sigma_z$ in 165 mock samples of Ly$\alpha$
emitters, the mean and 1 sigma dispersion are indicated by thick and
thin dashed lines respectively. In the mocks, Ly$\alpha$ emitters
are assumed to reside in halos more massive than $3\times
10^{11}\msun$. Full lines and filled squares neglect peculiar
velocities. The horizontal dotted line denotes the mean dispersion in
the absence of peculiar velocities and clustering. Halo clustering
decreases $\sigma_z$ significantly (dotted line compared to full line)
but peculiar velocities do not have a strong effect (dashed line
compared to full line). The observed points fall well within the range
covered by the mocks.}
\label{fig:clumping}
\end{figure}

To assess the likelihood of the observed $\sigma_z$ values and the
influence of peculiar velocities we extract 15 mock catalogues from
each of the P2 runs.  Each mock catalogue is extracted by picking
random redshift-space volumes with sizes corresponding to the volume
sampled by the observation and selecting all DM halos more massive
than $3\times10^{11}$ M$_\odot$ contained in the volume.  The
connection between minimal Ly$\alpha$ flux and minimal halo mass is
fixed loosely (see Sect.~4.1-4.3), so the number of emitters here is
to be considered as indicative.  However, as long as such small halos
trace the same structure nearly independently of mass, $\sigma_z$
should not be affected by this assumption.  Referring to a filter FWHM
of 60 \AA\ and a field of view of 6.7 arcmin, we extract volumes of
12.4 $\times$ 12.4 $\times$ 47.0 co-moving Mpc (the line of sight
corresponding to the longer dimension).  Boxes are required to contain
at least three objects.  Redshifts are computed along the major axis
of the extracted volume.  Fig.~\ref{fig:clumping} shows the resulting
$\sigma_z$ of the redshift distributions of the mock catalogues as a
function of the number of mock emitters found in the box which is a
measure of overdensity. The $\sigma_z$ values are computed both
neglecting and considering peculiar velocities. The lines show the
average and $\pm$1-$\sigma$ intervals of the $\sigma_z$
distribution. The expected $\sigma_z$ value in the case of no clumping is
0.0142; due to the well-known clustering of halos, significantly lower
values are expected on average.

The observational points are reported as well.  As the filters are
more similar to Gaussians than to top-hats, the expected $\sigma_z$
value for a uniform distribution (0.019) is higher than in our case
which assumed a top-hat (0.014), therefore we multiply the observed
values by $0.014/0.019=0.737$.

As it is apparent, peculiar velocities are responsible for decreasing
the value of $\sigma_z$ by some 10\% when it is already small; these
are cases of filaments (or pancakes) seen perpendicularly to the line
of sight, where the effect of flattening by peculiar velocities is
largest. The two observed points are well within the predicted range,
so these fields are by no means rare cases.  In particular, the low
value of $\sigma_z$ in the Q2138-4427 field, coupled to the moderately
high value of the overdensity inferred, can be interpreted, as
mentioned above, as the effect of a flattened structure. It is a
1.76$\sigma$ event so equally low values of $\sigma_z$ will be
expected in 4.5\% of all observed fields. If peculiar velocities are
neglected, the Q2138-4427 field turns out to be a 1.93$\sigma$ event,
only marginally rarer.

Pentericci et al.\ (2000) and Venemans et al.\ (2002) both use the
observed redshift clumping to argue for substantial overdensities
around radio galaxies, and claim these suggest the detection of a
protocluster. They assume that the overdensity $\delta$ can be
estimated from
\begin{equation}
\delta = \frac{n_{obs} \times \mathrm{fwhm}_{\mathrm{filter}}}
{n_{field} \times \mathrm{fwhm}_z},
\end{equation}
where $n_{obs}$ is the observed number density in the field containing
the radio galaxy, $n_{field}$ is an estimate of the number density of
a reference field and $\mathrm{fwhm}_{\mathrm{filter}}$ and
$\mathrm{fwhm}_z$ are the full-width-at-half-maxima of the filter
transmission (transformed to redshift space) and the observed redshift
distribution, respectively. However, as seen from
Fig.~\ref{fig:clumping} $\sigma_z$ is not a decreasing function of
density. In fact, it is more likely to have a low $\sigma_z$ in the
redshift distribution in a field with few Ly$\alpha$ emitters than in
an overdense field. Therefore, the only valid way of resolving whether
radio galaxies are located in protoclusters is to obtain an accurate
measurement of the number density of galaxies in blank fields at
similar redshifts.

\section{Conclusions}\label{summary}

We have characterised and quantified the effect of peculiar velocities
in the reconstruction of large-scale structure at high redshift, with
particular attention to Ly$\alpha$ emitters as tracers.

With the aid of \pin\, simulations we have decomposed the velocity
field of DM halos into a streaming flow, a gradient and a random
velocity term, and computed them as functions of scale.  The
dependence of these velocity statistics on halo mass, redshift and
cosmology has been quantified and fitting formulae have been proposed.

The main effects of these velocity components on the observational
properties of Ly$\alpha$ emitters have been analysed.  In particular,
streaming flows are determined by fluctuations on very large scales,
and are strongly affected by sample variance, but have a modest impact
on the interpretation of observations.  Gradient flows are mostly
important, in that they influence the quantitative reconstruction of
structures like the inclination angle of filaments, important for
applying the extended Alcock-Paczy\'nski test (M{\o}ller \& Fynbo
2001), or the root-mean-square of the redshift distribution, important
to recognise flattened structures (pancakes or filaments)
perpendicular to the line of sight.  Random velocities are typically
below or of the same order as the observational uncertainty on the
redshift.

The results presented here have been applied to quantify the influence
of peculiar velocity on the reconstructed viewing angles of filaments
at $z\simeq3$.  In particular, the effect of streaming velocities is
negligible, gradient velocities give an error of 2-3$^o$ degrees,
similar but larger than the typical error due to sparse sampling,
while random velocities add to the $\sim100$ km s$^{-1}$ error on the
redshift.  Clearly, a proper quantification of such errors is
necessary to implement an Alcock-Paczy\'nski test to the inclination
of filaments.

As a further example of the power of this approach, we have generated
mock catalogues of Ly$\alpha$ emitters to assess the significance of a
detected narrow distribution in redshift in a deep exposure. The
observation is found to be a $\sim$2$\sigma$ event corresponding to a
sheet of galaxies seen face on. Peculiar velocities give a modest but
significant contribution to the narrowness of the redshift
distribution, and this again corresponds to the dominant effect of
gradient velocities with respect to random velocities. Moreover, we do not notice
a significant anti-correlation between the abundance of emitters, a
tracer of overdensity, and the degree of clumpiness, at variance with
what is assumed by Pentericci et al.\ (2000) and Venemans et al.\
(2002).

The results presented here will be important for interpreting the
upcoming data on the large-scale structure as traced by Ly$\alpha$
emitters. Further work will be aimed at generating mock catalogues of
Ly$\alpha$ emitters that closely reproduce the observational selection
effects, in order to devise tight observational tests for the
hierarchical clustering model at $z\ga2$.

\section*{Acknowledgements}
We thank Stefano Borgani for making his simulation available.
P. Monaco thanks ESO for hospitality and support.  This work was
supported by the Danish Natural Science Research Council (SNF) and by
the Carlsberg Foundation. TT thanks PPARC for the award of an Advanced
Fellowship.  \pin\ can be downloaded from
http://www.daut.univ.trieste.it/pinocchio/.

\appendix
\section{Adding long wavelength modes to \pin}
Tormen and Bertschinger (1996) describe an algorithm to increase the
dynamic range of a simulation by adding long-wavelength perturbations
after the simulation has been done. However, as pointed out by Cole
(1997), the algorithm neglects the coupling between long-wavelength
linear modes and short-wavelength non-linear modes, and this strongly
affects the clustering of halos. Fortunately, this is not a problem in
\pin, since it is easy to correctly incorporate the effect of long wavelength
modes on the non-linear collapse of structures. We begin by
giving a very brief overview of the \pin\ algorithm, and then proceed
to describe how one can easily add long wavelength modes.

The standard \pin\ algorithm operates on a realisation of a
linear density field generated on a regular grid, identical to the
grid used in the initial conditions of an $N$-body simulation. In a
first step, a \lq collapse time\rq\ is computed for each grid point
(\lq particle\rq) using a truncation of Lagrangian perturbation theory
based on ellipsoidal collapse.  The collapse time is the time at which
the particle is deemed to fall into a high-density region (a halo or
filament). In the second step, collapsed particles are gathered into
halos, using an algorithm that mimics the hierarchical build-up of
halos (see Monaco et al.\ 2002a for more details).

The calculation of the collapse times itself also involves two steps,
(a) a series of linear operations on the initial density field,
followed by (b) a non-linear calculation. For a Gaussian random field,
the long- and short-wavelength perturbations are by definition
independent, therefore it is trivial to perform the first step for long
and short wavelengths separately. In contrast to the Tormen \&
Bertschinger (1996) implementation, the result of the calculation of
the two step procedure (i.e. doing long and short wavelengths separately)
gives identical result to doing the full calculation, yet requires
significantly less computation.

The algorithm works as follows. Take the linear potential
$\psi({\bf q})$, defined on the vertices ${\bf q}$ of a grid. The grid
spacing $\Delta$, together with the extent of the grid, $L$, determine
the range of waves that can be represented, namely between $2\Delta$
and $L$. However, consider now two grids, with spacings $\Delta_1$ and
$\Delta_2$, and extents $L_1$ and $L_2$ respectively. Grid 2
represents a higher resolution grid contained within grid 1, and we
want to add the long-wavelength perturbations of grid 1 onto grid 2,
increasing the dynamic range from $L_2/\Delta_2$ to $L_1/\Delta_2$.

On the vertices of grid 2, we can add the contributions from
fluctuations on grid 1 and grid 2 to obtain the potential $\psi$:

\begin{equation}
\psi({\bf q}) = \psi_1({\bf q}) + \psi_2({\bf q})\,,
\label{eq:psi}
\end{equation}

Clearly $\psi$ has contributions from the full range of waves,
$2\Delta_2$ to $L_1$. Of course the spacing of grid 1 is coarser than of
grid 2, $\Delta_1 > \Delta_2$, so equation~(\ref{eq:psi}) involves an
interpolation from the coarser to the finer grid. But the key point is
that, as long as the operations we are going to do on $\psi$ are
linear, we can perform them on grids 1 and 2 independently,
and just add the result at the end to compute the collapse time for the
vertices of the higher resolution grid. The rest of the \pin\,
calculation now only applies to the high resolution grid, but 
we have to be aware of boundary effects on the edge of the smaller
grid.

When initialising the Gaussian fluctuations on these grids, we use the
power spectrum $P(k)\,\Theta(k_1)$ on grid 1, and
$P(k)\,(1-\Theta(k_1))\,\Theta(k_2)$, where $P(k)$ is the desired
linear power-spectrum, and the Heaviside function restricts the
contribution from waves $>k_1$, respectively $k_2$.  $k_2$ denotes the
Nyquist frequency on the high-resolution grid, and $k_1$ should be
smaller than the Nyquist frequency of the lower-resolution grid but
larger than $2\pi/L_2$.

For the box and grid lengths given in
Section~\ref{section:simulations} ($L_2=65\, h^{-1}$,
$L_2/\Delta_2=256$, $L_1= 8 L_2=520\, h^{-1}$ Mpc, $L_1/\Delta_1=64$),
we found that a good choice for $k_1=\pi/L_1$. The effective dynamic
range of these simulations is thus $(L_1/\Delta_2)^3=2048^3$, whereas
the simulation time is more similar to performing two $256^3$
simulations. Given that the simulation time is dominated by the fast
Fourier transforms on the grid that scale as $N\log(N)$, with
$N=(L/\Delta)^3$, this is an acceleration of a factor of 352, and we 
effectively perform a $2048^3$ simulation in a few hours on a PC.

\begin{figure}
\centering{
\epsfig{file=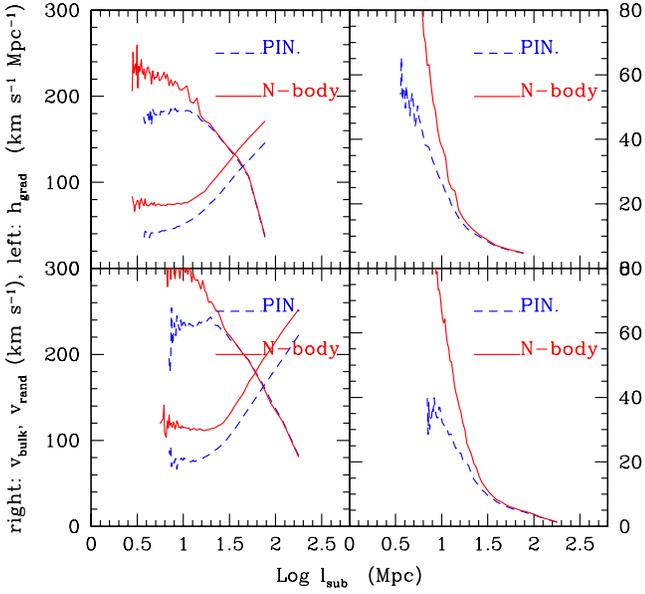,width=8.5cm}
}
\caption{Velocity statistics comparing \pin\, and $N$-body runs.  Upper
panels show the 100 Mpc/$h$ simulation of Monaco et al.\ (2002b) for
$M>3\times 10^{11}$ M$_\odot$ at $z=2$, lower panels the 250 Mpc/$h$
simulation of Fontanot et al.\ (2003) for $M>3\times 10^{12}$ M$_\odot$
at $z=0$}
\label{fig:nbody}
\end{figure}

\section{\pin\ accuracy in recovering peculiar velocity components}
To check the accuracy achieved by \pin\ in predicting the three
velocity statistics defined in Section~\ref{section:statistics} we
compare the \pin\ result to those of two different $256^3$ N-body
simulations, the 100Mpc/$h$ run used by Monaco et al.\ (2002b) and
Taffoni et al.\ (2002), and the 250 Mpc/$h$ presented by Fontanot et
al.\ (2003).  In both cases we run \pin\ with a single grid, and on
the same initial conditions as the simulations. The assumed
cosmologies are similar to the reference one, except that $h=0.65$ in the
first simulation, and $\sigma_8=0.8$ in the second one.  For the first
simulation, the mass resolution is a factor of 3 lower, so that $3\times
10^{11}$ M$_\odot$ halos are the smallest reliable ones.  For the
second simulation the mass of the particle is $\sim10^{11}$ M$_\odot$,
so that only halos with $M \ga 5\times 10^{12}$ M$_\odot$ are
reliable.  In order to have a sufficient number of halos, we test this
simulation at $z=0$.  Halos in both simulations have been selected
with the usual friends-of-friends algorithm with a linking length 0.2
times the interparticle distance (Jenkins et al.\ 2001).

Fig.~\ref{fig:nbody} shows the streaming, gradient and random velocity
statistics for the PINOCCHIO and N-body runs.  On scales larger
than 10 co-moving Mpc at $z=2$ (or 20 at $z=0$) \pin\ systematically underestimates the
random velocity  by $\sim$30 km $s^{-1}$ while it
reproduces fairly well the streaming velocity. The gradient component is
underestimated at worst by $\sim$30\%.  At smaller scales these
underestimates are larger, but the qualitative behaviour is always
reproduced.

This level of agreement is expected, because \pin\ velocities are based
on the Zel'dovich (1970) approximation, which is known to reproduce
well the large-scale velocity field but to underestimate the
small-scale, highly non-linear velocities.  The latter are the result
of infall of halos onto neighbours.

As the scales of interest are those relative to the large-scale
structures observed (like filaments), roughly corresponding to the
cross-over of streaming and random velocities, we conclude that \pin\
is sufficiently accurate for our present purpose.


\begin{thebibliography}{99}

\bibitem{A98}Adelberger, K., Steidel, C. C., Giavalisco, M. et al.\ 1998, ApJ, 505, 18
\bibitem{A}Alcock, C. \& Paczy\'nski, B. 1979, Nature, 281, 358
\bibitem[Berlind et al.(2003)]{2003ApJ...593....1B} Berlind, A.~A., et al.\  2003, \apj, 593, 1 
\bibitem[Cole(1997)]{1997MNRAS.286...38C} Cole, S.\ 1997, \mnras, 286, 38 
\bibitem{D2003}Daddi, E., R{\"o}ttgering, H. J. A., Labb{\'e}, I. et al.\ 2003, ApJ, 588, 50
\bibitem{fontanot03}Fontanot, F., Monaco, P. \& Borgani, S. 2003, MNRAS, 341, 692
\bibitem{FMT2001}Fynbo, J. P. U., M\o ller P., Thomsen, B. 2001, A\&A, 374, 443
\bibitem{FMT2003}Fynbo, J. P. U., Ledoux, C., M\o ller P., Thomsen, B., Burud, I. 2003, A\&A, 407, 147
\bibitem{HH}Haehnelt, M. G., Natarajan, P. \& Rees, M. J. 1998, MNRAS, 300, 817
\bibitem[Hamana et al.(2003)]{2003MNRAS.343.1312H} Hamana, T., Kayo, I., Yoshida, N., Suto, Y., \& Jing, Y.~P.\ 2003, \mnras, 343, 1312 
\bibitem{1983ApJS...52...89H} Huchra, J., Davis, M., Latham, D., \& Tonry, J.\ 1983, ApJS, 52, 89
\bibitem{j}Jenkins A., Frenk C.S., White S.D.M., Colberg J.M., Cole S., Evrard A.E., Couchman H.M.P. \& Yoshida N. 2001, MNRAS, 321, 372
\bibitem{M04}Miley, G. K., Overzier, R. A., Tsvetanov, Z. I. et al.\ 2004, Nature, 427, 47
\bibitem{M}M\o ller, P. \& Fynbo, J.U. 2001, A\&A, 372, L57
\bibitem{M02a}Monaco, P., Theuns, T., Taffoni, G., Governato, F., Quinn, T., \& Stadel, J., 2002a, ApJ, 564, 8
\bibitem{M02b}Monaco, P., Theuns, T., \& Taffoni, G. 2002b, MNRAS, 331, 587
\bibitem[Peebles(1980)]{1980lssu.book.....P} Peebles, P.~J.~E.\ 1980, Research supported by the National Science Foundation.~Princeton, N.J., Princeton University Press, 1980.~435 p.,  
\bibitem{P00}Pentericci, L., Kurk, J. D., R{\"o}ttgering, H. J. A. et al.\ 2000, A\&AL, 361, 25
\bibitem[Persic et al.(1996)]{1996MNRAS.281...27P} Persic, M., Salucci, P., \& Stel, F.\ 1996, \mnras, 281, 27 
\bibitem{Sh}Shapley, A., et al.\ 2003, ApJ, 588, 65
\bibitem{S03}Shimasaku, K., Ouchi, M., Okamura, S. et al.\ 2003, ApJ, 586, L111
\bibitem[Spergel et al.(2003)]{2003ApJS..148..175S} Spergel, D.~N., et al.\ 2003, \apjs, 148, 175 
\bibitem{S00}Steidel, C. C., Adelberger, K. L., Shapley, A. E. et al.\ 2000, ApJ, 532, 170
\bibitem{Straus}Strauss, M. A. \& Willick, J. A. 1995, Phys Rep, 261, 271
\bibitem{T02}Taffoni, G., Monaco, P., \& Theuns, T. 2002, MNRAS, 333, 623
\bibitem[Tormen \& Bertschinger(1996)]{1996ApJ...472...14T} Tormen, G.~\& Bertschinger, E.\ 1996, \apj, 472, 14 
\bibitem{V03}Venemans, B. P., Kurk, J. D., Miley, G. K. et al.\ 2002, ApJL, 569, 11
\bibitem{WM96}Warren, S. J. \& M\o ller, P. 1996, A\&A, 311, 25
\bibitem{W02}Weidinger, M., M\o ller, P., Fynbo, J. P. U., Thomsen, B. \& Egholm, M. P. 2002, A\&A, 391, 13
\bibitem{z}Zel'dovich YA. B. 1970, Astrofizika, 6, 319 (translated in Astrophysics, 6, 164 [1973])
\end{thebibliography}
\end{document}